\documentclass[reprint,
superscriptaddress,
amsmath,amssymb,
aps,
prl,
floatfix,
]{revtex4-2}
\usepackage[colorlinks=true,citecolor=blue,linkcolor=red,urlcolor=blue]{hyperref}

\usepackage{microtype}

\usepackage{graphicx}

\newcommand{\mnras}{Mon.\ Not.\ R.\ Astron.\ Soc.}
\newcommand{\apjl}{Astrophys.\ J.}
\newcommand{\aap}{Astron.\ Astrophys.}
\newcommand{\araa}{ARA\&A}
\newcommand{\procspie}{Proc.\ SPIE}

\begin{document}

\title{High-Precision Determination of Oxygen-K\texorpdfstring{$\alpha$}{a} Transition Energy Excludes Incongruent Motion of Interstellar Oxygen}

\author{M.~A.~Leutenegger}\email{maurice.a.leutenegger@nasa.gov}
\affiliation{NASA Goddard Space Flight Center, 8800 Greenbelt Rd., Greenbelt, MD 20771, USA}
\author{S.~K\"uhn}
\affiliation{Max-Planck-Institut f\"ur Kernphysik, Saupfercheckweg 1,
  69117 Heidelberg, Germany}
\author{P.~Micke}
\affiliation{Max-Planck-Institut f\"ur Kernphysik, Saupfercheckweg 1,
  69117 Heidelberg, Germany}
\affiliation{Physikalisch-Technische Bundesanstalt, Bundesallee 100, 38116 Braunschweig, Germany}
\author{R.~Steinbr\"ugge}
\affiliation{Deutsches Elektronen-Synchrotron DESY, Notkestr.~85, 22607
  Hamburg, Germany}
\author{J.~Stierhof}
\affiliation{Remeis-Sternwarte \& Erlangen Centre for Astroparticle
  Physics, Friedrich-Alexander-Universit\"at Erlangen-N\"urnberg, Sternwartstr.~7, 96049 Bamberg, Germany}
\author{C.~Shah}
\affiliation{NASA Goddard Space Flight Center, 8800 Greenbelt Rd., Greenbelt, MD 20771, USA}
\affiliation{Max-Planck-Institut f\"ur Kernphysik, Saupfercheckweg 1,
  69117 Heidelberg, Germany}
\author{N.~Hell}
\affiliation{Lawrence Livermore National Laboratory, 7000 East Ave,
  Livermore, CA 94550, USA}
\author{M.~Bissinger}
\affiliation{Erlangen Centre for Astroparticle
  Physics, Friedrich-Alexander-Universit\"at Erlangen-N\"urnberg, Friedrich-Alexander-Universit\"at Erlangen-N\"urnberg, Erwin Rommel Str.~1, 91058 Erlangen, Germany}
\author{M.~Hirsch}
\affiliation{Remeis-Sternwarte \& Erlangen Centre for Astroparticle
  Physics, Friedrich-Alexander-Universit\"at Erlangen-N\"urnberg, Sternwartstr.~7, 96049 Bamberg, Germany}
\author{R.~Ballhausen}
\affiliation{Remeis-Sternwarte \& Erlangen Centre for Astroparticle
  Physics, Friedrich-Alexander-Universit\"at Erlangen-N\"urnberg, Sternwartstr.~7, 96049 Bamberg, Germany}
\author{M.~Lang}
\affiliation{Remeis-Sternwarte \& Erlangen Centre for Astroparticle
  Physics, Friedrich-Alexander-Universit\"at Erlangen-N\"urnberg, Sternwartstr.~7, 96049 Bamberg, Germany}
\author{C.~Gr\"afe}
\affiliation{Remeis-Sternwarte \& Erlangen Centre for Astroparticle
  Physics, Friedrich-Alexander-Universit\"at Erlangen-N\"urnberg, Sternwartstr.~7, 96049 Bamberg, Germany}
\author{S.~Wipf}
\affiliation{Institut f\"ur Optik und Quantenelektronik, Friedrich-Schiller-Universit\"at Jena, Max-Wien-Platz 1, 07743 Jena, Germany}
\author{R.~Cumbee}
\affiliation{NASA Goddard Space Flight Center, 8800 Greenbelt Rd., Greenbelt, MD 20771, USA}
\affiliation{Department of Astronomy, University of Maryland, College Park, MD 20742}
\author{G.~L.~Betancourt-Martinez}
\affiliation{Institut de Recherche en Astrophysique et Plan\'etologie, 9, avenue du Colonel Roche BP 44346, 31028 Toulouse Cedex 4, France}
\author{S.~Park}
\affiliation{Ulsan National Institute of Science and Technology, 50 UNIST-gil, Ulsan, South Korea}
\author{V.~A.~Yerokhin}
\affiliation{Peter the Great St.~Petersburg Polytechnic University, 195251 St.~Petersburg, Russia}
\author{A.~Surzhykov}
\affiliation{Physikalisch-Technische Bundesanstalt, Bundesallee 100, 38116 Braunschweig, Germany}
\affiliation{Institut f\"ur Mathematische Physik, Technische Universit\"at Braunschweig, D-38106 Braunschweig, Germany}
\author{W.~C.~Stolte}
\altaffiliation[Presently affiliated with Nevada National Security Site, managed and operated by ]{Mission Support and Test Services, Livermore operations, Livermore CA 94551}
\affiliation{Advanced Light Source, Lawrence Berkeley National Laboratory, Berkeley, CA 94720, USA}
\author{J.~Niskanen}
\affiliation{Institute for Methods and Instrumentation in Synchrotron Radiation Research G-ISRR, Helmholtz-Zentrum Berlin f\"ur Materialien und Energie, Albert-Einstein-Strasse 15, 12489 Berlin, Germany}
\affiliation{Department of Physics and Astronomy, University of Turku, FI-20014 Turun Yliopisto, Finland}
\author{M.~Chung}
\affiliation{Ulsan National Institute of Science and Technology, 50 UNIST-gil, Ulsan, South Korea}
\author{F.~S.~Porter}
\affiliation{NASA Goddard Space Flight Center, 8800 Greenbelt Rd., Greenbelt, MD 20771, USA}
\author{T.~St\"ohlker}
\affiliation{Institut f\"ur Optik und Quantenelektronik, Friedrich-Schiller-Universit\"at Jena, Max-Wien-Platz 1, 07743 Jena, Germany}
\affiliation{GSI Helmholtzzentrum f\"ur Schwerionenforschung, Planckstra{\ss}e 1, 64291 Darmstadt, Germany}
\affiliation{Helmholtz-Institut Jena, Fr\"obelstieg 3, 07743 Jena, Germany}
\author{T.~Pfeifer}
\affiliation{Max-Planck-Institut f\"ur Kernphysik, Saupfercheckweg 1,
  69117 Heidelberg, Germany}
\author{J.~Wilms}
\affiliation{Remeis-Sternwarte \& Erlangen Centre for Astroparticle
  Physics, Friedrich-Alexander-Universit\"at Erlangen-N\"urnberg, Sternwartstr.~7, 96049 Bamberg, Germany}
\author{G.~V. Brown}
\affiliation{Lawrence Livermore National Laboratory, 7000 East Ave,
  Livermore, CA 94550, USA}
\author{J.~R.~{Crespo L\'opez-Urrutia}}
\affiliation{Max-Planck-Institut f\"ur Kernphysik, Saupfercheckweg 1,
  69117 Heidelberg, Germany}
\author{S.~Bernitt}
\affiliation{Institut f\"ur Optik und Quantenelektronik, Friedrich-Schiller-Universit\"at Jena, Max-Wien-Platz 1, 07743 Jena, Germany}
\affiliation{Helmholtz-Institut Jena, Fr\"obelstieg 3, 07743 Jena, Germany}
\affiliation{GSI Helmholtzzentrum f\"ur Schwerionenforschung, Planckstra{\ss}e 1, 64291 Darmstadt, Germany}
\affiliation{Max-Planck-Institut f\"ur Kernphysik, Saupfercheckweg 1,
  69117 Heidelberg, Germany}

\date{\today}

\begin{abstract}

We demonstrate a widely applicable technique to absolutely calibrate the energy scale of x-ray spectra with experimentally well-known and accurately calculable transitions of highly charged ions, allowing us to measure the K-shell Rydberg spectrum of molecular O$_2$ with 8\,meV-uncertainty. We reveal a systematic $\sim$450\,meV shift from previous literature values, and settle an extraordinary discrepancy between astrophysical and laboratory measurements of neutral atomic oxygen, the latter being calibrated against the aforementioned O$_2$ literature values. Because of the widespread use of such, now deprecated, references, our method impacts on many branches of x-ray absorption spectroscopy. Moreover, it potentially reduces absolute uncertainties there to below the meV level.
\end{abstract}

\maketitle

The vast majority of baryonic matter in the Universe appears as diffuse gas at temperatures ranging from 10\,K to 10\,MK \cite{2012ApJ...759...23S}. Owing to the presence of elements heavier than hydrogen and helium, which have strong inner-shell absorption features in the 0.2--2\,keV band, x-ray observations provide a sensitive means to trace this gas and to determine its properties \cite{2002ApJ...573..157N, 2012ApJ...753...17C, 2014MNRAS.442.3745M, 2017ARA&A..55..389T}. As oxygen is the third most abundant element in the Universe \cite{lodders2020}, the strong $1s\mbox{ -- }2p$ resonance line from atomic oxygen is especially important for such studies. Its strength provides a measure of the abundance, and its Doppler shift yields the radial velocity.

\begin{figure*}
  \includegraphics[width=0.82\textwidth]{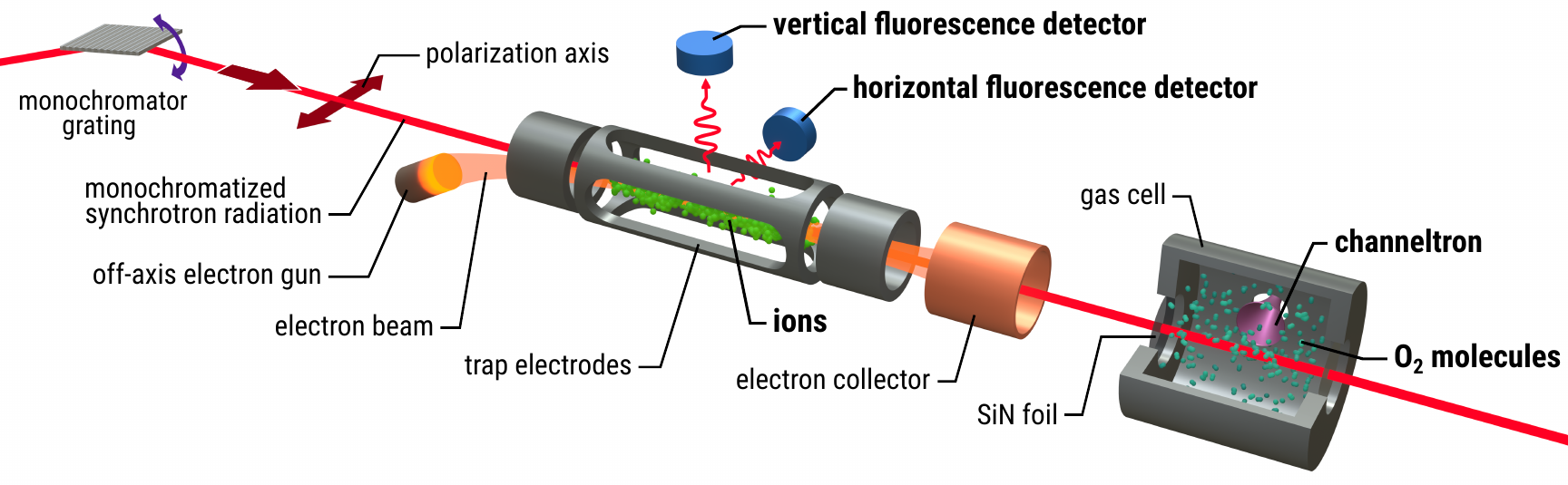}
  \caption{Scheme for simultaneous measurements of resonant fluorescence spectra of highly charged ions trapped in a compact electron beam ion trap equipped with an off-axis electron gun \cite{2018RScI...89f3109M} and photoabsorption spectra of molecular gases in a separated gas cell downstream. A monochromatic synchrotron radiation beam, which is slowly scanned in energy, passes through both the ion trap and the gas cell, thus eliminating the effect of temporal drifts and providing a stable mutual reference for the spectra of ionized and neutral species.}
  \label{fig:expsetup}
\end{figure*}

To enable this science, space instruments such as the \textit{Chandra} High Energy Transmission Grating Spectrometer (HETGS) have been calibrated to better than $100\,\mathrm{km}\,\mathrm{s}^{-1}$ \cite{2004SPIE.5165..457M, 2006ApJ...644L.117I}. The stability of this calibration has been tracked on-orbit through repeated observations of soft x-ray transitions from highly charged ions (HCI), specifically of H- and He-like ions of elements such as neon, oxygen, and nitrogen, in the coronae of stars with small and known radial velocities \citep[e.g.,][]{Cel00, 2006ApJ...649..979G} or of supernova remnants \citep[e.g.,][]{plucinsky:2012}, and verified through observations with other space instruments such as the \textit{XMM-Newton} Reflection Grating Spectrometer (RGS; \citep{2015A&A...573A.128D}) and x-ray sensitive CCDs \citep{plucinsky:2017}. For the $1s\mbox{ -- }2p$ oxygen resonance line, the HETGS has yielded radial velocity measurements with an uncertainty as low as $13\,\mathrm{km}\,\mathrm{s}^{-1}$ \cite{2013ApJ...768...60G}. Surprisingly, \citet{2013ApJ...779...78G} showed that averaging measurements of this line over different lines of sight in the Galaxy did not yield a value close to the rest value as was expected. The average wavelength differed from the best laboratory value \cite{1997JPhB...30.4489S, 2013ApJ...771L...8M, 2015PhRvA..92b3401B} by an amount equivalent to ${\sim}340\,\mathrm{km}\,\mathrm{s}^{-1}$, i.e., outside the laboratory and HETG uncertainties. RGS data also appear to require a shift of  ${\sim}380\,\mathrm{km}\,\mathrm{s}^{-1}$, when compared with theory calibrated against the same laboratory measurements \citep{devries:2003}. To put this shift in context, the Galactic escape velocity in the vicinity of the solar system is $(580\pm63)\,\mathrm{km}\,\mathrm{s}^{-1}$ \citep{monari:2018}.

The traditional calibration standard for the laboratory measurements of the atomic oxygen spectrum was the conveniently measurable absorption spectrum of molecular oxygen. Its value was established by electron energy loss spectroscopy (EELS) measurements \cite{1974JElSpR..4..313W, 1980JElSpR..18..21H}. Conversely, the absolute wavelength calibration of the grating spectrometers of \textit{Chandra} and \textit{XMM-Newton} outlined above primarily relies on soft x-ray transitions from H-like and He-like ions of elements such as neon, oxygen, and nitrogen in objects with well known radial velocities.

\begin{figure}[b]
  \includegraphics[width=\columnwidth]{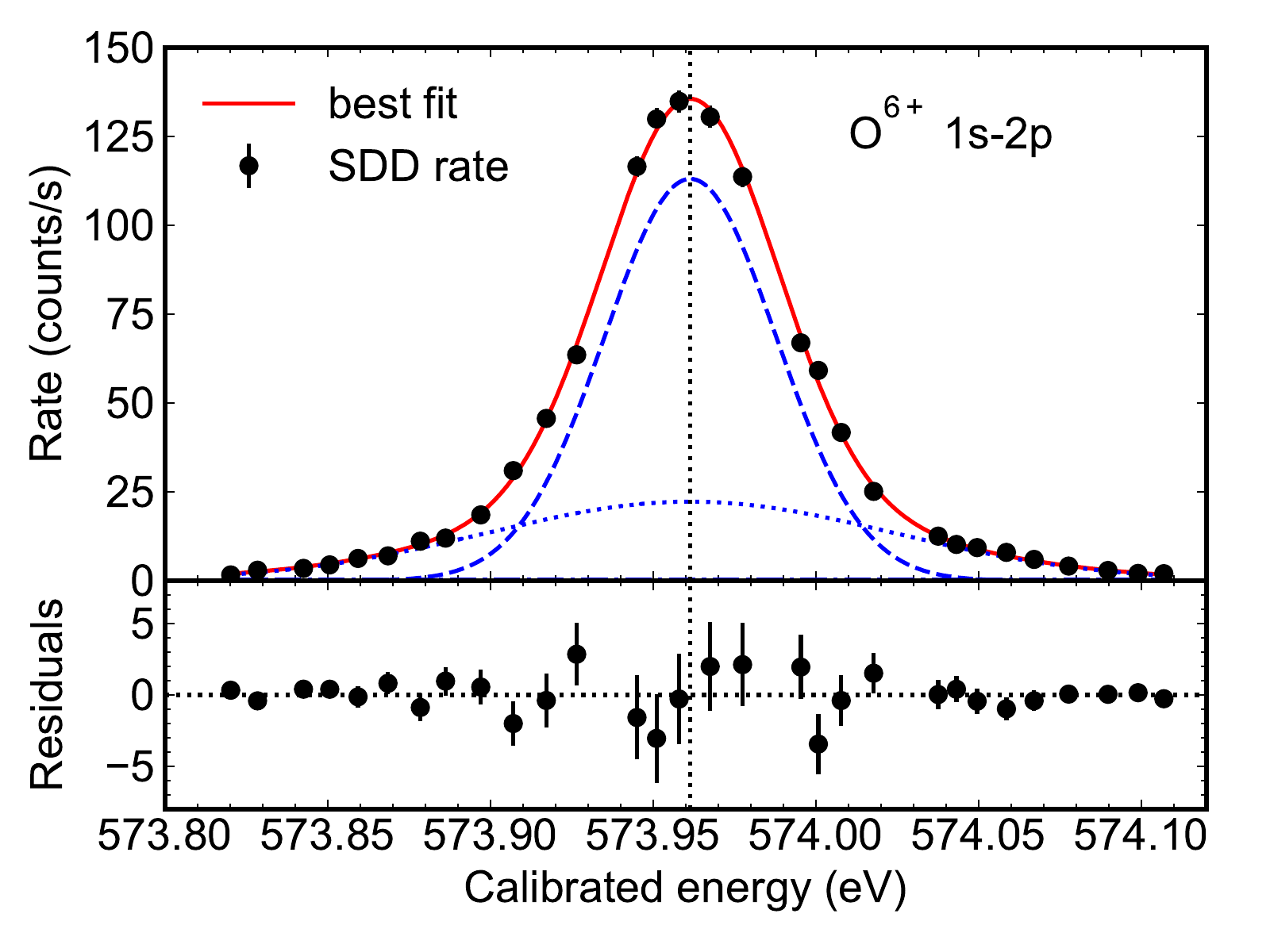} 
  \caption{Example of a calibration scan. We recorded $\sim$ 15000 counts in the vertical SDD in a 15-minute scan. The best-fit model is the sum of two co-centered Gaussians, reflecting a non-ideal instrumental line shape, and has a FWHM of 69 meV. The centroid of the $1s\mbox{ -- }2p$ resonance transition of He-like $\mathrm{O}^{6+}$ has a statistical uncertainty of 0.3\,meV, while the theoretical uncertainty is 0.53\,meV.} \label{fig:owexample}
\end{figure}

In this Letter, we introduce an independent, accurate laboratory calibration technique in order to help resolve this puzzling and significant discrepancy between space-based and laboratory energy calibration methods. We measure the molecular oxygen Rydberg spectrum simultaneously with x-ray lines from He-like ions; specifically, we present a new high-precision measurement of the K-shell absorption spectrum of molecular oxygen using the well-known $1s \mbox{ -- } np$ resonance transitions (i.e. $1s^2\, ^1S_0 \mbox{ -- } 1s\, np\, ^1P_1$) of He-like $\mathrm{O}^{6+}$ and $\mathrm{N}^{5+}$ as calibration references. This reduces the uncertainty of the laboratory standard to only $4\,\mathrm{km}\,\mathrm{s}^{-1}$, unveils a significant calibration error in the hitherto used standard, and brings the laboratory energy scale into agreement with the calibration of space-based instruments. Our method overcomes current limitations and outperforms the accuracy of existing soft x-ray calibration standards by at least three orders of magnitude. 

Our setup (Fig.~\ref{fig:expsetup}) was installed at beamline U49-2/PGM-1 \cite{osti_20652951, kachel2016} of the synchrotron-radiation facility BESSY-II (Helmholtz-Zentrum Berlin), where an undulator delivers linearly polarized light to a plane-grating monochromator, with typical photon fluxes of $10^{12}\,\mathrm{s}^{-1}$ in the energy range of 500--600\,eV. An exit slit width of 10\,$\mu$m yielded a full-width-at-half-maximum (FWHM) resolution of 69\,meV (E/$\Delta$E = 8320) at 574\,eV.

\begin{figure*}
\includegraphics[width=\textwidth]{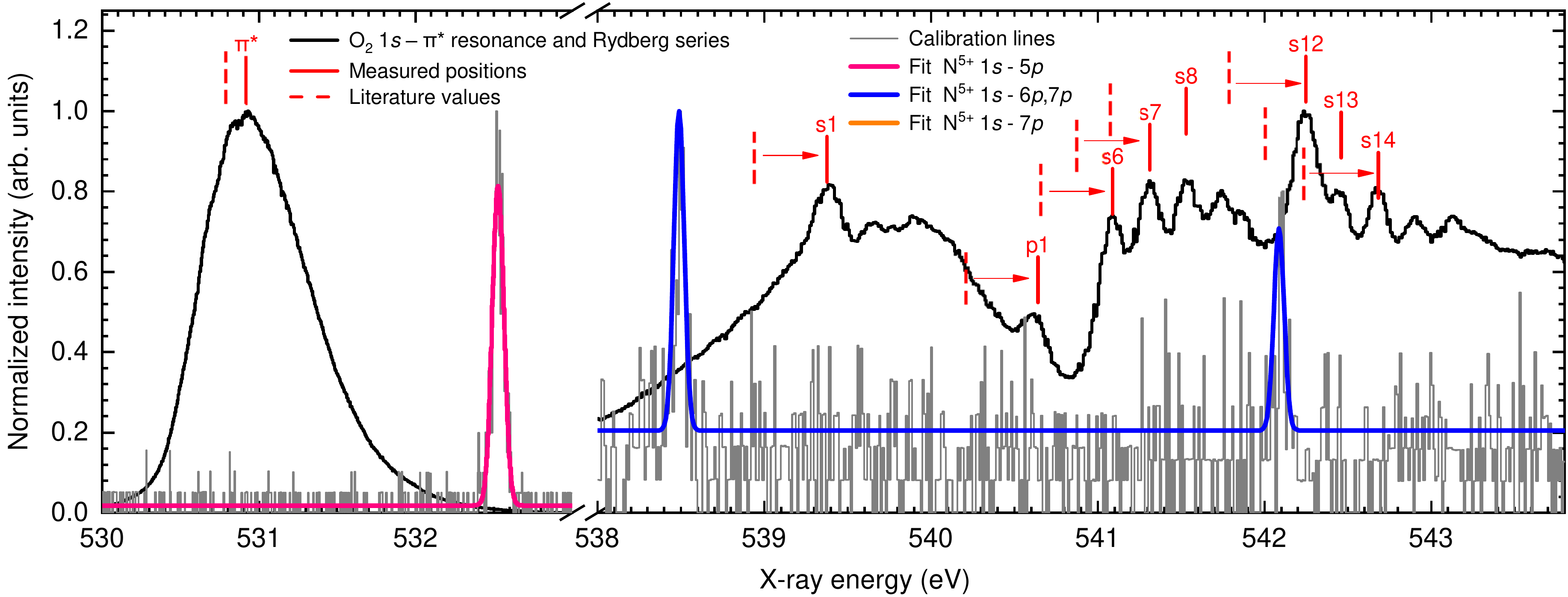} 
\caption{Recalibration of the O$_2$ soft x-ray absorption spectrum by simultaneously measuring $K_\delta$ ($1s\mbox{ -- }5p$), $K_\epsilon$ ($1s\mbox{ -- }6p$), and $K_\zeta$ ($1s\mbox{ -- }7p$) transitions in He-like N$^{5+}$ as energy references. Positions of spectral features in O$_2$ from the literature \cite{1974JElSpR..4..313W, 2008PhRvA..78b2516T} (dashed red vertical markers), are compared with our measurements (full red vertical markers), clearly showing the energy offset (see Tab.~\ref{tab:O2ref}). The vertical scales are different for the two spectral regions shown.}
\label{fig:o2_recalibration}
\end{figure*}

We used PolarX-EBIT \cite{2018RScI...89f3109M}, an electron beam ion trap (EBIT) employing a novel off-axis electron gun leaving the main axis obstacle-free, to produce and store HCI by means of its monoenergetic electron beam. The photon beam merges on the longitudinal axis of PolarX-EBIT with the electron beam \cite{2007PhRvL..98r3001E, 2012Natur.492..225B, 2013PhRvL.111j3002R}, and passes through the device to the absorption cell downstream. An electron beam at an energy of 420\,eV (300\,eV), well below the K-shell excitation energies of $\mathrm{O}^{6+}$ ($\mathrm{N}^{5+}$), produces and traps the ions for study from injected N$_2$ and residual H$_2$O that are dissociated by it. These choices of electron-beam energy suppress excitation of soft x-ray transitions in the energy band of interest both by electron impact and resonant as well as non-resonant photorecombination processes. A low beam current of only ${\sim}1.1$\,mA reduces ion heating and its associated Doppler broadening, as originally shown by \citet{BeiersdorferPRL1996}. 

Fluorescence photons from the decay of photoexcited ions were detected with two (one vertical, the other horizontal) silicon drift detectors (SDDs) mounted side-on to the photon beam axis. Photon events were recorded with a multi-channel data acquisition (DAQ) system. Since the photon beam was horizontally polarized, the signal was stronger in the vertical detector for $J=0 \mbox{ -- } 1$ transitions~\cite{balashovbook}. This effect is most pronounced for the $1s \mbox{ -- } 2p$ transitions and decreases for $1s \mbox{ -- } np$ transitions with higher principal quantum number $n$ due to depolarization effects in alternate decay paths.  The transition energy does not depend on polarization.

Two meters downstream from PolarX-EBIT, a cell continuously fed with O$_2$ gas using a needle valve and pumped down to keep a constant pressure of ${\sim}10^{-6}$\,mbar is installed (Fig.~\ref{fig:expsetup}). A 30 nm SiN foil separated the vacua of the gas cell and PolarX-EBIT. For detection of single photoions produced by absorption of the soft x-rays in the gas, we used a channeltron. Its electronic pulses were amplified, pulse-height discriminated, and passed to the DAQ system. Background was negligible when the x-ray beam was off; while on, but without O$_2$ gas injection, residual gaseous H$_2$O (at ${\sim}10^{-7}$\,mbar) leads to weak O K-shell transitions cleanly distinguishible from that of O$_2$ \cite{WIGHT197425}.

The photon energy is selected by rotations of the monochromator plane grating and its ancillary mirror that are measured with high resolution encoders. Regular calibrations are needed because of thermal drifts in the positions of beamline optical components, encoder errors, and shifts in the x-ray source position caused by adjustments to the storage ring orbit parameters. We slowly scanned the photon energy across the ranges of interest; at each position, our DAQ system recorded the grating and mirror angles, nominal photon energy, counts from SDDs and channeltrons, and storage ring current.

We show an example calibration scan of the $1s\mbox{ -- } 2p$ transition of He-like O$^{6+}$ in Fig.~\ref{fig:owexample}. Within a 15-minute-long measurement, it achieved a statistical uncertainty for the centroid position of 0.3\,meV, smaller than the theoretical uncertainty of the transition energy and better than 1\,ppm in precision. We performed similar scans of the $1s\mbox{ -- } np$ transitions of O$^{6+}$ and N$^{5+}$ up to $n=7$.

Repeated scans displayed drifts of the photon energy on the order of 50--300\,meV on timescales of several hours to days. Consecutive scans of the same line sometimes showed drifts of 10--30\,meV within 50\,minutes. Slow, smooth shifts while scanning over a 1 eV scan region of the O$_2$ Rydberg spectrum in the gas cell did not exceed 40 meV; thus we assume a systematic uncertainty proportional (40 meV per eV) to the separation from the nearest $1s\mbox{ -- }np$ calibration line of N$^{5+}$ (see Supplemental Material).

We therefore simultaneously calibrated the O$_2$ $1s\mbox{ -- }\pi^*$ transition and Rydberg series with the $1s\mbox{ -- }5p$, $1s\mbox{ -- }6p$, and $1s\mbox{ -- }7p$ transitions of N$^{5+}$ in the same broad scans. We used transition energies for N$^{5+}$ from \citet{doi:10.1063/1.5121413}, which are calculated with techniques \cite{PhysRevA.81.022507, PhysRevA.96.042505} that have been experimentally benchmarked to 1.5\,ppm for $1s\mbox{ -- }2p$ transitions in He-like Ar \cite{2014PhRvA..90c2508K}. The theoretical uncertainty for the $1s\mbox{ -- }5p,\,6p,\,7p$ transition energies is estimated to be 0.3\,meV \cite{doi:10.1063/1.5121413}. Our recalibrated spectra are shown in Fig.~\ref{fig:o2_recalibration}, together with best fit peak positions and previously published reference positions \cite{2008PhRvA..78b2516T} tracing their calibration to the original EELS measurements\cite{1974JElSpR..4..313W, 1980JElSpR..18..21H}. 

\begin{table}
  \caption{Energies measured for selected peaks in the O$_2$ Rydberg series compared with measurements from \citet{2008PhRvA..78b2516T} (labeled T08 below). Peak labels and assignments are as in that work. Uncertainties of T08 are relative, and do not include the absolute error of the calibration standard used.\label{tab:O2ref}}
  \begin{ruledtabular}
    \begin{tabular}{clllcc}
      Peak & \multicolumn{3}{c}{Energy (eV)} & \multicolumn{2}{c}{Assignment} \\
           & \multicolumn{1}{c}{This work} & \multicolumn{1}{c}{T08} & \multicolumn{1}{c}{Shift} & $^4\Sigma^-$ & $^2\Sigma^-$ \\
      \hline
      s1 & 539.377(35) & 538.95(4) & 0.427 & $3s\sigma$ $\nu\!=\!0$ & \\
      p1 & 540.641(58) & 540.22(4) & 0.421 & $3p\pi$ & \\
      s6 & 541.089(40) & 540.67(4) & 0.419 & $3p\sigma$ $\nu\!=\!0$ &  \\
      s7 & 541.313(31) & 540.89(4) & 0.423 & $3p\sigma$ $\nu\!=\!1$ & \\
      s8 & 541.530(22) & 541.09(4) & 0.440 & $3p\sigma$ $\nu\!=\!2$ & \\
      s12 & 542.249(8) & 541.80(4) & 0.449 & $4p\sigma$ $\nu\!=\!0$ & $3p^\prime \sigma$ $\nu\!=\!0$\\
      s13 & 542.459(19) & 542.02(5) & 0.439 & $4p\sigma$ $\nu\!=\!1$ & $3p^\prime \sigma$ $\nu\!=\!1$ \\
      s14 & 542.683(24) & 542.25(5) & 0.433 & 4$d\sigma$ & \\
    \end{tabular}
  \end{ruledtabular}
\end{table}

To get the best possible calibration for the strongest feature (s12) in the O$_2$ Rydberg series, we used the $1s\mbox{ -- }7p$ resonance of $\mathrm{N}^{5+}$, separated from it by only 158\,meV, with a calculated energy of 542.09057(31)\,eV. The peaks appear in the scan only 4\,minutes apart, avoiding systematic shifts affecting longer timescales. We derived a peak energy of 542.249(8)\,eV, with an uncertainty of 5\,meV from counting statistics on the N$^{5+}$ $1s\mbox{ -- }7p$ transition, an estimated 3\,meV systematic contribution from the fits of neighboring peaks, and 6\,meV from drift due to the 158\,meV separation from N$^{5+}$ $1s\mbox{ -- }7p$. We assign a larger 12\,meV uncertainty to the nearest peak at 542.459(19)\,eV (s13) to account for its greater sensitivity to the fit model of the dominant s12 peak, in addition to the 15\, meV drift uncertainty. For all other peaks (Tab.~\ref{tab:O2ref}), uncertainties are dominated by those of the drift; all peaks are referenced to N$^{5+}$ $1s\mbox{ -- }7p$, with the exception of $\pi_*$ (N$^{5+}$ $1s\mbox{ -- }5p$) and s1 (N$^{5+}$ $1s\mbox{ -- }6p$).

Our result for peak s12 differs by 0.449\,eV (-248~km~s$^{-1}$) from the value that was originally measured by~\citet{1980JElSpR..18..21H} and that has been used as a standard in numerous works \citep[e.g., ][]{KOSUGI1992481, PhysRevLett.72.3961, 2008PhRvA..78b2516T}, including for atomic oxygen \cite{1997JPhB...30.4489S, 2013ApJ...771L...8M}. We find similar shifts for the rest of the O$_2$ Rydberg series. However, the peak of the $1s\mbox{ -- }\pi^*$ transition, which is calibrated with respect to the N$^{5+}$ $1s\mbox{ -- }5p$ transition, is measured to be 530.92(6)\,eV, which is shifted by only 0.12\,eV from the value reported by \citet{1974JElSpR..4..313W}, well within their quoted uncertainty of 0.2\,eV. It is not clear why the shift in the calibration of the O$_2$ Rydberg series is 0.33\,eV larger, since \citet{1980JElSpR..18..21H} referenced the Rydberg series against $1s\mbox{ -- }\pi^*$. Real peak shifts of the temperature-dependent rovibrational distribution in $1s\mbox{ -- }\pi^*$, and the quoted 0.1\,eV energy uncertainties might explain this.

We recalibrated the data set of \citet{2013ApJ...771L...8M} using our measured energy for the strongest Rydberg peak (s12) at 542.249(8)\,eV. The fitting uncertainty for this peak was 7\,meV, yielding a net calibration uncertainty of 11\,meV. We then performed a new fit of the nearby $1s\mbox{ -- }3p\,^4P$ line of atomic oxygen, which had an 8\,meV fit uncertainty, yielding a total uncertainty of 14\,meV. Finally, we fitted the $1s\mbox{ -- }2p$ line of atomic oxygen, obtaining a best fit value of 527.26(4)\,eV. Here the uncertainty is dominated by scan-to-scan calibration shifts across the 14.4\,eV separating $1s\mbox{ -- }2p$ and $1s\mbox{ -- }3p\,^4P$.

\begin{table}[t]
  \caption{Recalibrated energies for atomic oxygen (in eV) compared with previous works (G13: \citet{2013ApJ...779...78G}; M13: \citet{2013ApJ...771L...8M}; L13: \citet{2013ApJ...774..116L}). Doppler shifts relative to our work given in $\mathrm{km}\,\mathrm{s}^{-1}$; sh.\ and ave.\ refer to shifted and averaged (see text). The XMM results are for Mkn 421, while the Chandra results are for a weighted average of multiple lines of sight. \label{tab:Oref}}
  \begin{ruledtabular}
    \begin{tabular}{llrlr}
      Source & $1s\mbox{ -- }2p$ & \multicolumn{1}{c}{$\Delta v$} & $1s\mbox{ -- }3p\,{}^4P$ & \multicolumn{1}{c}{$\Delta v$} \\
      \hline
      This work & 527.26(4) &  & 541.645(12) & \\ 
      \hline
      G13 \textit{XMM} & 527.28(5) & $-11$(36) & 541.93(28) & $-158$(155) \\
      G13 \textit{XMM}, sh. & 527.30(5) & $-22$(36) & 541.95(28) & $-169$(155) \\
      G13 \textit{Chandra}  & 527.44(9) & $-102$(56) & 541.72(18) & $-42$(100) \\
      G13 \textit{Ch.}, sh. & 527.26(9) & $-11$(56) & & \\     
      L13 \textit{Chandra}& 527.39(2) & $-74$(25) \\ 
      \hline 
      M13 ALS & 526.79(4) & 267(32) & 541.19(4) & 252(22) \\
    \end{tabular}
  \end{ruledtabular}
\end{table}

Our recalibrated line energies for $1s\mbox{ -- }2p$ and $1s\mbox{ -- }3p\,{}^4P$ in neutral oxygen are much closer to previously published astrophysical values for neutral gas in the intergalactic medium, as shown in Tab.~\ref{tab:Oref}. Independently from each other, \citet{2013ApJ...779...78G} and \citet{2013ApJ...774..116L} averaged \textit{Chandra} spectra for multiple lines of sight in the galaxy. \citet{2013ApJ...779...78G} also analyze a high signal-to-noise \textit{XMM-Newton} RGS spectrum of Mkn 421. In Table~\ref{tab:Oref} we show both the values of the line position determined using the instrumental wavelength calibration and the values as corrected by \citeauthor{2013ApJ...779...78G} based on observed shifts of the O~\textsc{vii} line relative to the most precise laboratory measurements  \citep[21.60195\,\AA, 573.949\,eV, ][]{1995JPhB...28.2565E}. Our results for $1s\mbox{ -- }2p$ disagree with the \textit{Chandra} averages at the level of $0.13 \mbox{ --  }0.18$\,eV, corresponding to a velocity of $- 75 \mbox{ -- } 100\,\mathrm{km}\,\mathrm{s}^{-1}$, and agree with the Mkn 421 value from the RGS within uncertainties. This disagreement may reflect some combination of real astrophysical velocities such as motion of the absorbers with respect to the Galactic rotation, or residual calibration uncertainties. Indeed, \citet{2013ApJ...779...78G} indicate that in the case of XTE J1817$-$330, observations of the O~\textsc{vii} $1s \mbox{ -- } 2p$ line by \citet{2013ApJ...768...60G} are shifted by $\sim$9\,m\AA\ with respect to laboratory measurements \citep{1995JPhB...28.2565E}. Correcting for this shift, \citet{2013ApJ...779...78G} find a line energy of 527.26(9)\,eV, fully consistent with the laboratory value found here. A more advanced description of the O~\textsc{vii} line as a blend of absorption and emission components by \citet{2013ApJ...774..116L} gives a quantitatively similar result for the shift. Taken together with our new calibration, this implies an astrophysical origin of this 9\,m\AA\ shift, corresponding to a velocity of $115\,\mathrm{km}\,\mathrm{s}^{-1}$, which is larger than expected from either the barycentric correction of the satellite's motion or from the rotational velocity of the Galaxy on the line of sight towards XTE J1817$-$330, and therefore may suggest an association of the absorber with this X-ray binary.

There is a growing need for reliable, easily reproducible energy calibration references over the whole x-ray band at modern high-flux radiation sources of steadily improving resolution and stability. Advanced synchrotron-radiation sources \cite[e.g.,][]{Bilderback2005} and free-electron lasers \cite{Feldhaus2005,Boutet2018} serve many x-ray absorption and scattering applications in biology, materials science, physical chemistry, as well as condensed-matter, atomic and molecular physics \cite{MinoRMP2018}. Subtle chemical, isotopic and crystallographic x-ray absorption shifts are studied in a plethora of x-ray absorption near-edge structure (XANES), extended as well as near-edge x-ray absorption fine structure (EXAFS, NEXAFS) experiments \cite{Stoehr1992} and with sophisticated theory \cite{RehrRMP2000}. Future radiation sources based on high-harmonic generation \cite{Couprie2014} will also require accurate photon-energy references. 

Calibration based on EELS suffers, i.~a., from systematic effects in the measurement of voltages applied to macroscopic electrodes. In view of the present results, it can be assumed that some unknown or underestimated uncertainties were extant but were not included by \citet{1980JElSpR..18..21H}, and were not corrected since then.
Other sophisticated methods to determine the dispersion function of grating or crystal spectrometers are hindered by natural limitations of the measurement techniques for distances, angles, grating spacing, and crystal lattice constants \cite{Becker1994,Ferroglio2008,Massa2009}. Furthermore, all these input parameters are sensitive to thermal shifts and mechanical vibrations.
The K-shell and L-shell lines of neutral atoms, widely used as x-ray energy standards \cite{DelattesRMP2003},  suffer from the presence of multiple blended satellite transitions that cannot be calculated with the high accuracy now possible for few-electron ions. They also display asymmetric line profiles affected by chemical and solid-state effects. Absorption edges used for calibration \cite{KraftRSI1996} are broader than those, and show even larger susceptibility to environmental influences.

In contrast, x-ray fluorescence lines in HCI are symmetric \cite{2013PhRvL.111j3002R,2012Natur.492..225B}, and can be, by choice of their multipolarity, as narrow as necessary for a given application. Their transition probabilities and level lifetimes span many orders of magnitude, and their energies are far more stable than other standards. This is true under all for our device conceivable values of temperature and electron density -- e.~g., extrapolating from \cite{GuPRA2020} for an electron-density effect on K$_{\beta}$ ($1s\mbox{ -- }3p$) of He-like Cl in an EBIT yields a shift lower than 1\,neV -- and recommends them as inherently superior references.
Furthermore, since space observatories often use naturally occurring HCI transitions for calibration, comparing them with the identical ones from an EBIT is straightforward. This can help when transitions from other isoelectronic sequences (e.~g., in \cite{Schlesser2013}) with larger theoretical uncertainties than the He- and H-like systems are investigated. 

The here introduced method is the most accurate presently available, being based on \emph{ab initio} calculations of He-like systems that have become extremely reliable during the last decades \cite[e.g.,][]{1988CaJPh..66..586D, PlanteJohnsonSapirstein1994, ArtemyevPRA2005}, with uncertainties reduced to a level well below 1\,meV in the recent work of \citet{doi:10.1063/1.5121413}. Hydrogenic transitions \cite{1985ADNDT..33..405J,Flowers2004}, which could in principle also be used with our method, reach below the part-per-billion uncertainty level, basically only limited by uncertainties on the nuclear size parameters. Following an analogous approach to optical frequency metrology with atoms and ions, x-ray energy references based on HCI can become ideal tools not only for calibration, but also for fundamental physics studies \cite{KozlovRMP2018} relying on exquisitely accurate measurements of photon energies and their shifts. Fully exploiting this technique, however, requires long-term stability of the experimental setup to a level of 1\,K in temperature and microns in mechanical stability, which are achievable in current state of the art facilities.  

\section{Acknowledgements}

\begin{acknowledgments}

Financial support was provided by the Max-Planck-Gesellschaft (MPG) and Bundesministerium f\"ur Bildung und Forschung (BMBF) through project 05K13SJ2. We thank HZB for the allocation of synchrotron radiation beamtime at BESSY II. Work by C.S.\ was supported by the Deutsche Forschungsgemeinschaft (DFG) Project No.\ 266229290 and by an appointment to the NASA Postdoctoral Program at the NASA Goddard Space Flight Center, administered by Universities Space Research Association under contract with NASA. Work by UNIST was supported by the National Research Foundation of Korea (No.\ NRF-2016R1A5A1013277). Work by LLNL was performed under the auspices of the U. S. Department of Energy under Contract No.\ DE-AC52-07NA27344 and supported by NASA grants to LLNL. M.A.L.\ and F.S.P.\ acknowledge support from NASA's Astrophysics Program. Work by G.B. was supported by a NASA Space Technology Research Fellowship. Work by R.C. was supported by an appointment to the NASA Postdoctoral Program at the NASA Goddard Space Flight Center.

\end{acknowledgments}

\bibliographystyle{apsrev4-2}
\bibliography{Master}

\end{document}


\title{High-Precision Determination of Oxygen-K\texorpdfstring{$\alpha$}{a} Transition Energy Excludes Incongruent Motion of Interstellar Oxygen: Supplemental Material}

\maketitle

\section{Evaluation of the BESSY II U49-2/PGM-1 energy scale calibration stability}

We evaluate the stability of the energy scale calibration of BESSY II beamline U49-2/PGM-1 in four ways: we consider how stable the energy scale is from one scan point to the next during a single scan by examining jitter in the flux of the $1s-3p$ transition of atomic Ne gas measured in the gas cell; we compare consecutive scans of $1s\mbox{ -- }np$ transitions in He-like ions; we compare non-consecutive scans of $1s\mbox{ -- }np$ transitions in He-like ions separated by an intervening scan of a different energy range; and we compare multiple scans of the O$_2$ Rydberg series to test the relative stability of the energy scale over single extended scans spanning $\sim$ 2 eV.

\subsection{Point-to-point jitter}
\label{sec:jitter}

\begin{figure}[!ht]
  \includegraphics[width=\columnwidth]{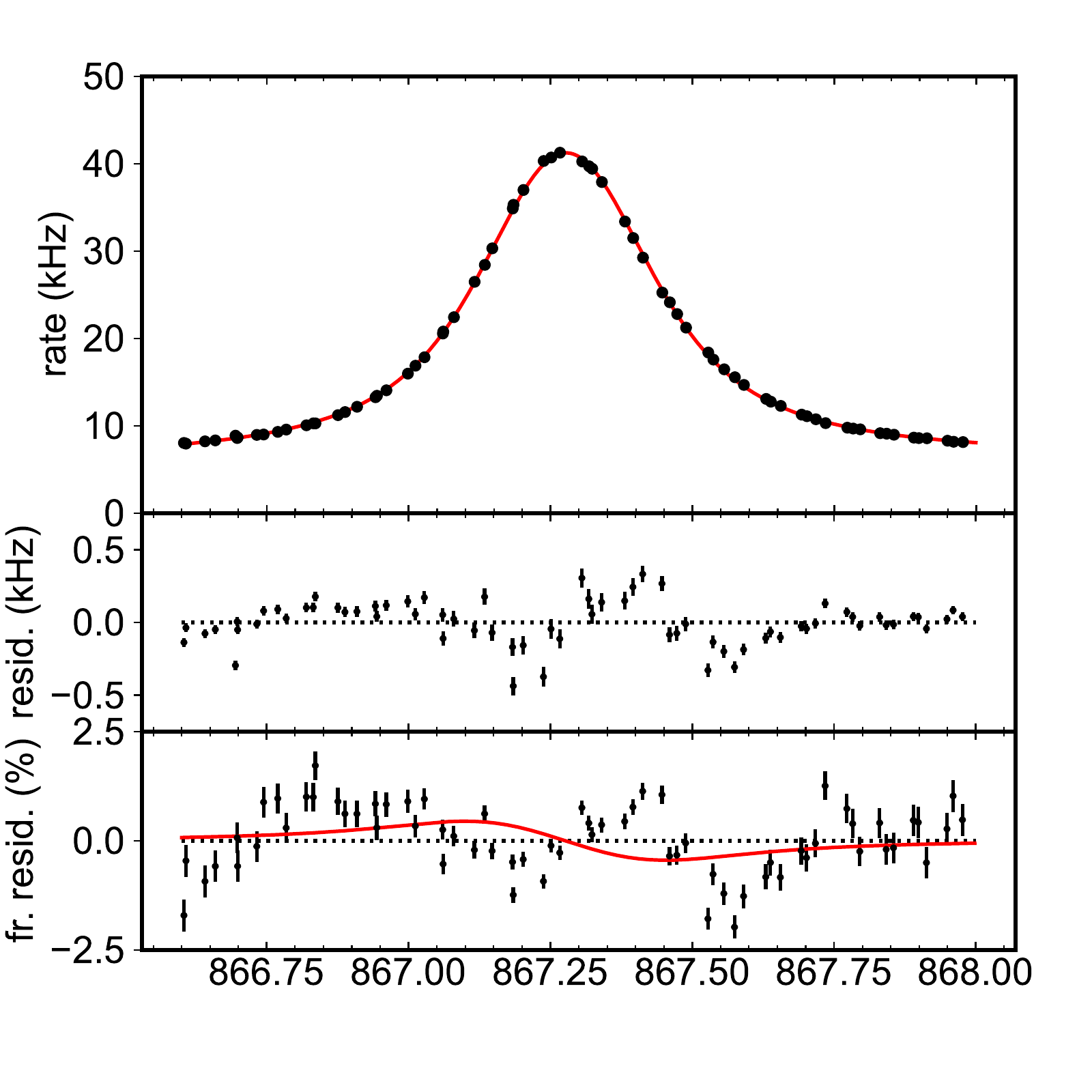}
  \caption{$1s\mbox{ -- }3p$ transition of atomic Ne in the gas cell, with the data and model shown in the top panel, the absolute residuals in the middle panel, and percent residuals shown in the bottom panel. Solid black circles with error bars represent the data. The solid red curve in the top panel shows the best-fit Voigt plus linear model. The solid red curve in the bottom panel is the derivative of the best-fit model times 1 meV, and corresponds to the flux jitter amplitude expected for an energy scale jitter of 1 meV. } \label{fig:Ne_gas_cell}
\end{figure}

We scanned the $1s\mbox{ -- }3p$ transition of atomic Ne in the gas cell, and we fit it with a Voigt model added to a linear term to account for background from residual gases and other terms in the Ne absorption cross section, as shown in Figure~\ref{fig:Ne_gas_cell}. In the bottom panel we show the fractional residuals as a percentage of the model value. We also show the product of the derivative of the model multiplied by 1 meV; this corresponds to the amplitude of point-to-point flux variability that would be expected if there were a 1 meV jitter in the energy compared with the reported energy. The error bars on each point correspond to the uncertainties from counting statistics only. While there are non-negligible residuals, the shape of the residuals as a function of energy indicates that there is a small systematic error in the model of the peak shape. The scatter of the points in the energy range where the derivative is highest is consistent with an energy calibration jitter of no more than 1 meV. Note that other effects, such as intrinsic fluctuations in the source flux, could contribute to flux jitter, and the 1 meV energy calibration jitter is thus an upper limit.

\subsection{Repeatability of consecutive scans}
\label{sec:consecutive}

\begin{figure}[!ht]
  \includegraphics[width=\columnwidth]{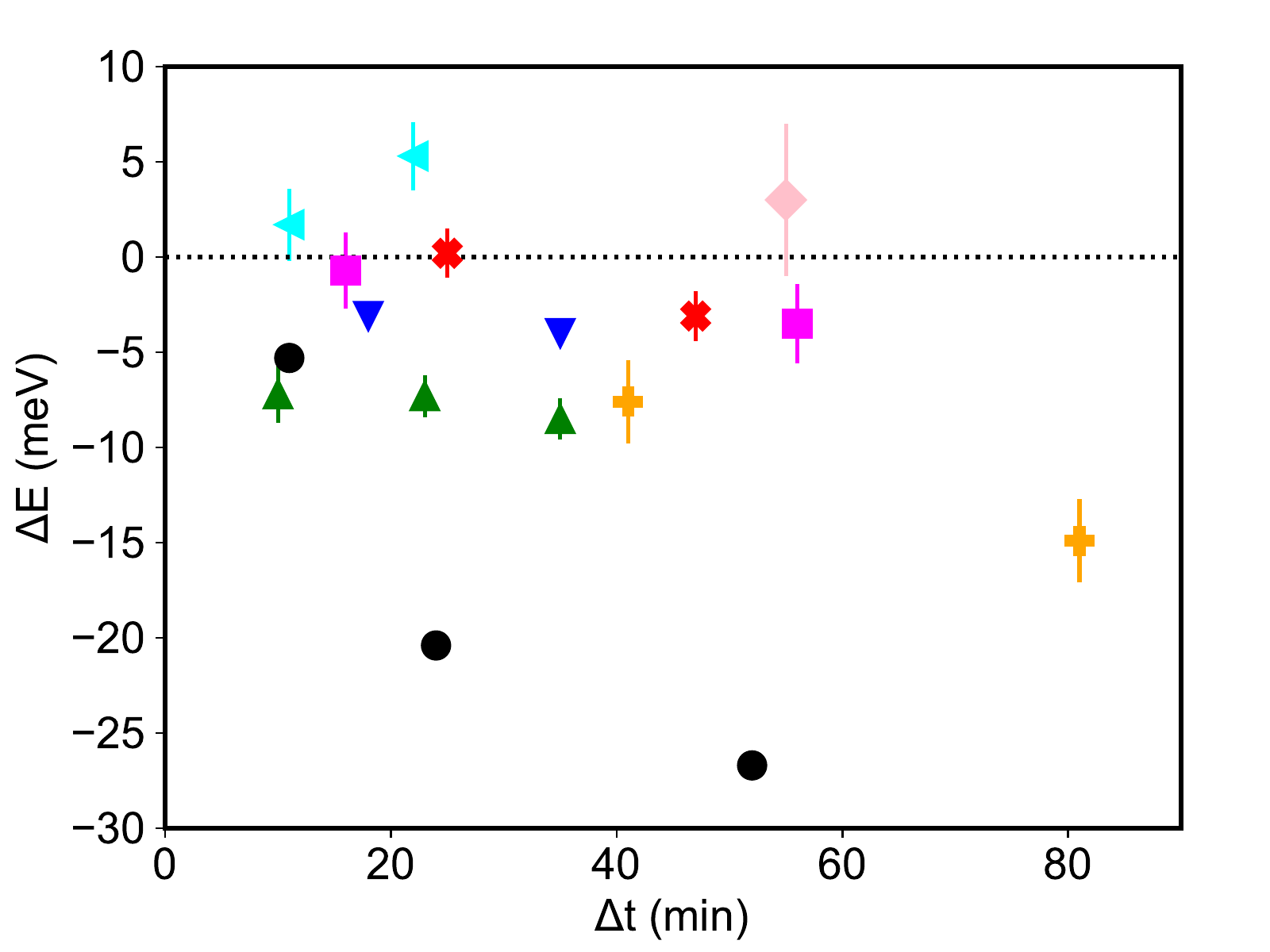}
  \caption{Relative shifts of repeated consecutive centroid measurements of the same transition as a function of lag after the first measurement in the series. Different symbols refer to different transitions: black circles are O$^{6+}\, 1s\mbox{ -- }2p$; magenta squares are O$^{5+}\, q\,  (1s\mbox{ -- }2p)$; red Xes are O$^{6+}\, 1s\mbox{ -- }3p$; orange plus signs are O$^{6+}\, 1s\mbox{ -- }4p$; pink diamonds are O$^{6+}\, 1s\mbox{ -- }5p$; blue downward pointing triangles are N$^{5+}\, 1s\mbox{ -- }2p$; green upward pointing triangles are N$^{5+}\, 1s\mbox{ -- }3p$; and cyan left pointing triangles are N$^{5+}\, 1s\mbox{ -- }4p$.} \label{fig:shifts_consecutive}
\end{figure}

To evaluate the stability of the beamline energy scale on timescales of tens of minutes and when not moving the monochromator optics to very different positions, we compared consecutive scans of the same $1s\mbox{ -- }np$ lines in N$^{5+}$, O$^{6+}$, or O$^{5+}$. In Figure~\ref{fig:shifts_consecutive} we show the centroid shift of each measurement in the series relative to the first measurement as a function of the time lag after the first measurement. We find typical shifts of order $\pm$ 5 - 10 meV, with a largest shift of -27 meV after 52 minutes.

\subsection{Repeatability of non-consecutive scans}
\label{sec:nonconsecutive}

\begin{figure}[!ht]
  \includegraphics[width=\columnwidth]{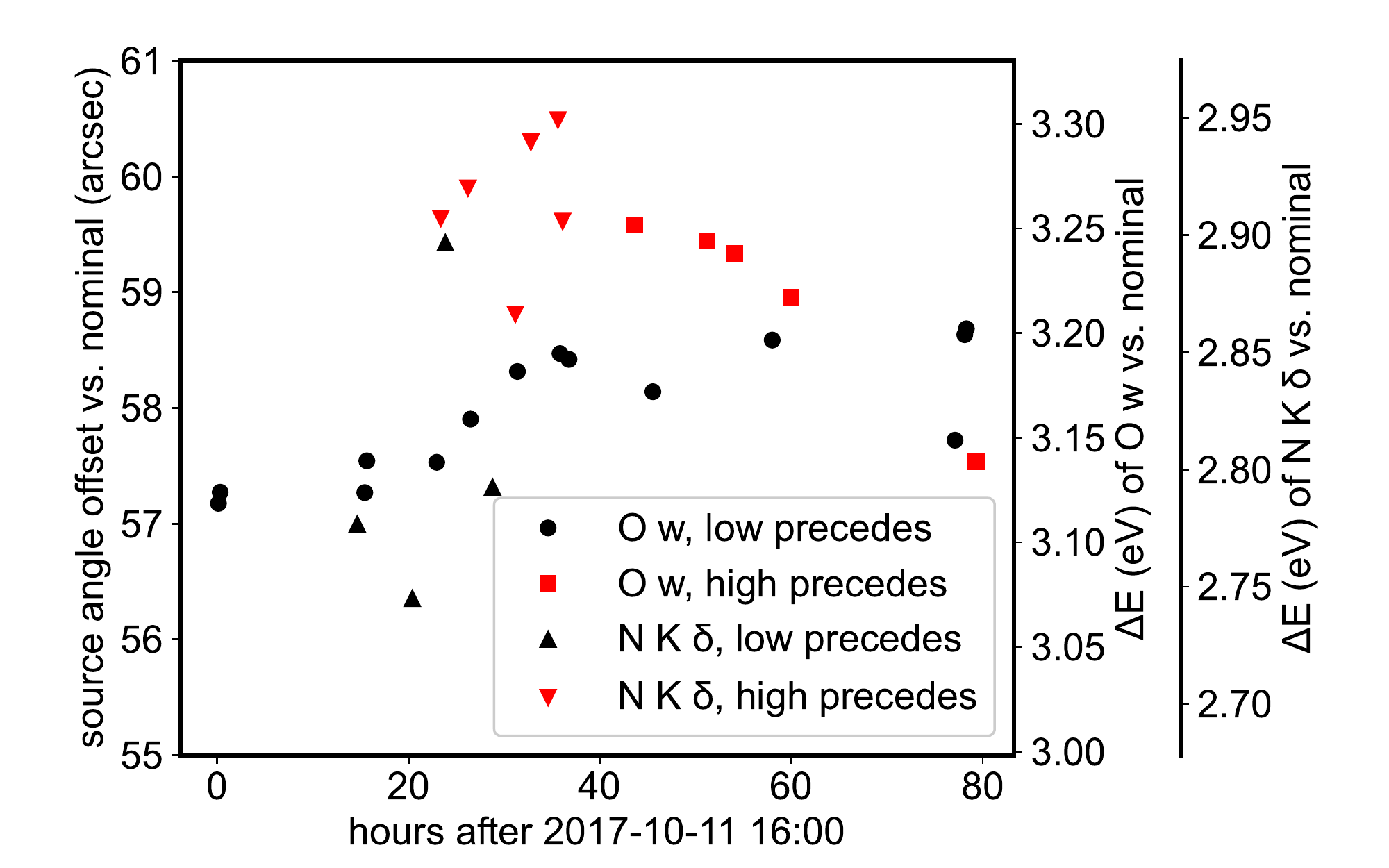}
  \caption{Offset in angle relative to nominal (left y-axis) and offset in calibrated energy relative to nominal (right y-axes) for N$^{5+}\, 1s\mbox{ -- }5p$ and O$^{6+}\, 1s\mbox{ -- }2p$, shown as a function of time during the experiment campaign. (Note that the large offset of $\sim$ 3 eV in the {\it nominal} energy scale is a known issue for this beamline, and that recalibrating against a known standard is routine.) The data points are also segregated into two groups depending on the state of the monochromator immediately preceding the measurement: "high precedes" refers to scans preceded by monochromator scans at a higher energy, while "low precedes" are preceded by monochromator scans at a lower energy.} \label{fig:ow_angles}
\end{figure}

To test the effect of larger motions of the monochromator, we evaluated the energy shifts of many scans of the $1s\mbox{ -- }5p$ transition of N$^{5+}$ and the $1s\mbox{ -- }2p$ transition of O$^{6+}$ throughout our experimental campaign. We show these measurements in Figure~\ref{fig:ow_angles}. We express the measured line positions as an energy shift relative to the nominal energy of the beamline (right y-axes), and as an angular shift in the source, equivalent to a shift in the incident angle $\alpha$ on the grating relative to the nominal angle (left y-axis)\footnote{Note that at a single energy point a shift in energy can be expressed equally well as a shift in source position, exit slit position, or a combination of the two. We have chosen in Figure~\ref{fig:ow_angles} to follow the convention of a pure source angle displacement}. These are reported as a function of time during the experiment campaign, and the points are grouped by whether a lower or higher energy scan precedes the measurement. We found that scans preceded by a higher energy tended to be offset to higher correction angles relative to those preceded by a lower energy. We speculate that this could be due to a hysteresis effect in the motions of the monochromator, or that it could be due to a time-variable heatload effect in the monochromator optics. A more careful and systematic study would be required to distinguish between these possibilities.

\subsection{Drift during scans over a broad energy range}
\label{sec:drift}

\begin{figure}[t]
  \includegraphics[width=\columnwidth]{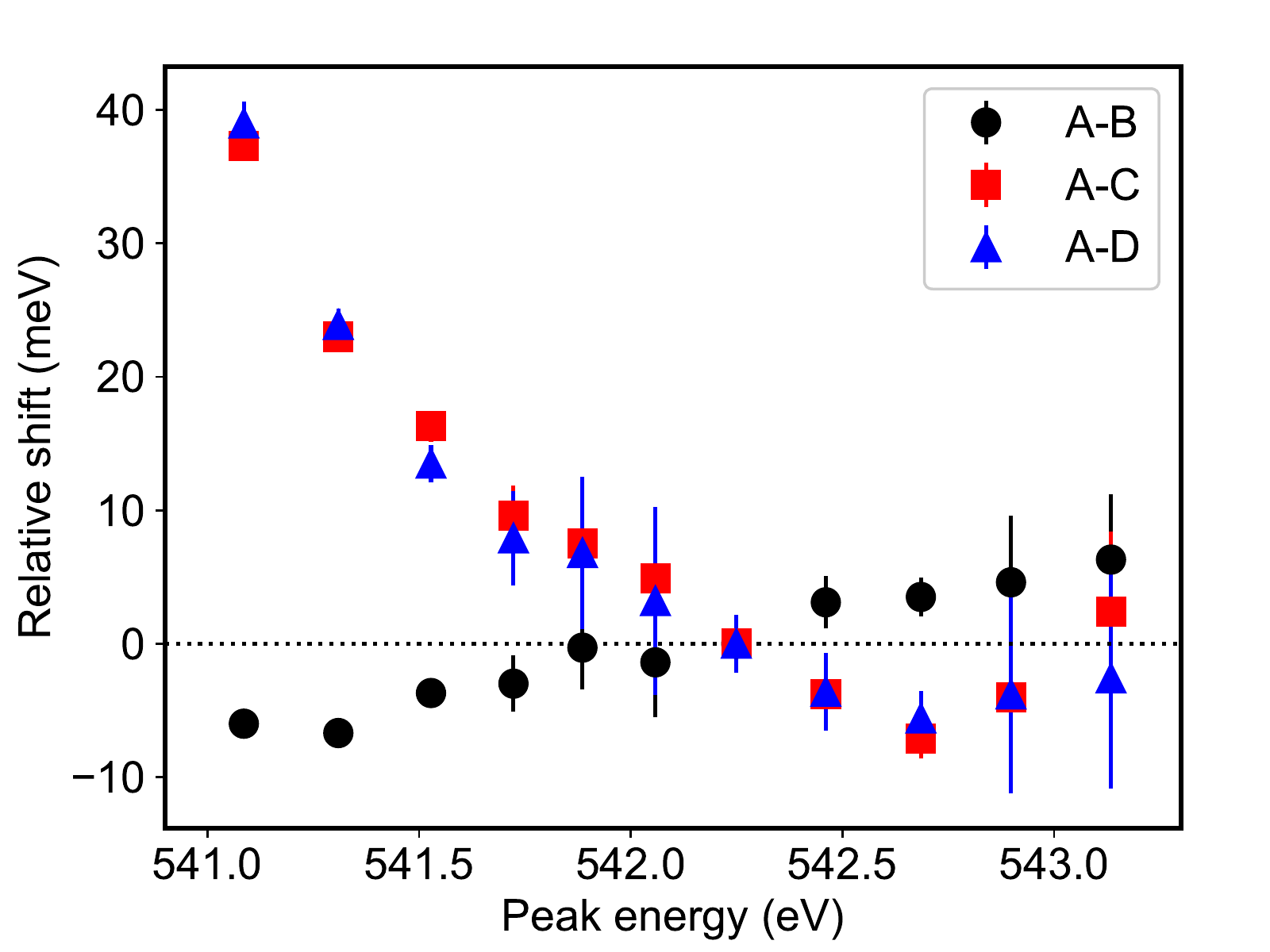}
  \caption{Relative shift of lines in the O$_2$ Rydberg series for four scans (labeled A-D). All scans are calibrated to set peak s12 to 542.249 eV, the value derived in this work. Scans A and B were both preceded by scans at lower energies, while scans C and D were both preceded by scans at higher energies. We speculate that the pattern of relative shifts may be due to a differential thermal relaxation effect.} \label{fig:compare_Rydberg}
\end{figure}

We tested the stability of the energy scale during scans over a broad energy range ($\sim$  2 eV) by comparing repeated scans of the O$_2$ Rydberg series, labeled A-D in chronological order, as shown in Figure~\ref{fig:compare_Rydberg}. Scans A and B were used together with the $1s\mbox{ -- }7p$ line of N$^{5+}$ to produce the absolute measurements of the Rydberg series reported in this article. During scans C and D the EBIT did not produce lines from N$^{7+}$ because of an issue with the sample injection system, so we can only use scans C and D by calibrating them against scans A and B. We thus used peak s12 as a reference for all four scans, and assessed the drift by comparing the relative shifts of the other peaks in the Rydberg series. 

We found that scans A and B agree quite well over the whole Rydberg series, with no shifts larger than $\pm$~10~meV. On the other hand, scans C and D have a larger offset of almost 40~meV at the beginning of the scan (near 541 eV). Scans C and D were immediately preceded by scans at higher energies, while scans A and B were immediately preceded by scans at lower energies. Given the results of Section~\ref{sec:nonconsecutive}, we speculate that the same thermal or hysteresis effect is present here, and that the effect gradually relaxes over the course of the scan.

\begin{figure*}[ht]
  \includegraphics[width=\textwidth]{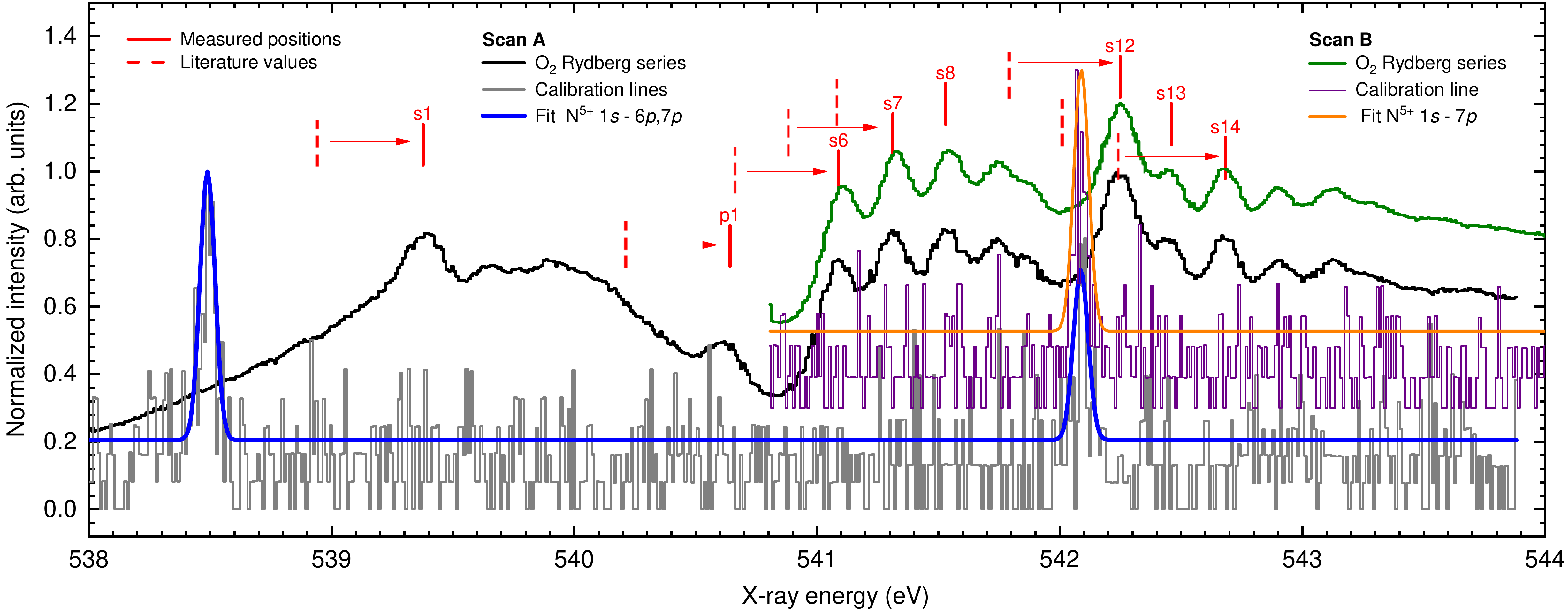}
  \caption{O$_2$ Rydberg series, showing energy range where two calibration scans were performed. \label{fig:both_scans}}
\end{figure*}

\subsection{Conclusions regarding energy scale stability}

Based on the measurements shown in Sections~\ref{sec:jitter}-\ref{sec:drift}, we reach the following conclusions: first, large energy shifts of $\sim$ tens of meV are usually associated with large relative motions of the monochromator energy from scan to scan; second, shifts or distorations during a scan are typically not large, with possible shifts up to 40 meV per eV, which may however be associated with differences in the state of the monochromator in the preceding measurement. We therefore assign a systematic uncertainty to each peak in the Rydberg spectrum of O$_2$ of 40 meV per eV of shift relative to the closest calibration line of N$^{6+}$.

\section{Comparison of O$_2$ Rydberg series scan data}

For clarity, Figure 3 of the main article shows only the Rydberg series data from scan A. In Figure~\ref{fig:both_scans} we show both scans A and B over the energy range where they overlap. 

The two scans were aligned using peak s12, and the $1s\mbox{ -- }7p$ transition of N$^{5+}$ was fit jointly for the two data sets, fixing the absolute energy scale calibration. We used this technique instead of calibrating each scan on its own best fit value of N$^{5+}$ $1s \mbox{ -- } 7p$ because of the relatively smaller statistical uncertainty on centroid positions in the O$_2$ Rydberg series.